\RequirePackage[hyphens]{url} 
\documentclass[%
 reprint,
 superscriptaddress,
 showpacs,amsmath,amssymb,
 aps,
floatfix,
nofootinbib
]{revtex4-1}
\usepackage{url}
\usepackage[breaklinks]{hyperref}
\PassOptionsToPackage{hyphens}{url}\usepackage{hyperref}
\usepackage{mathpazo}
\usepackage{times}
\usepackage{graphicx}

\usepackage{dcolumn}%
\usepackage{footnote}
\usepackage{makecell,multirow}
\usepackage{booktabs}
\usepackage[utf8]{inputenc}
\usepackage{multirow}
\usepackage{siunitx}
\usepackage{amsmath}
\usepackage{bm}
\usepackage[mathlines]{lineno}
\usepackage{braket}
\usepackage{color,soul}
\usepackage{subfigure}
\usepackage{changepage}
\usepackage[bottom]{footmisc}

\newsavebox{\foobox}

\begin{document}
\preprint{APS/123-QED}

\title{Dark Matter Axion Search Using a Josephson Traveling Wave Parametric Amplifier}
\author{C.Bartram}\email[Correspondence to: ]{chelsb89@uw.edu}
\author{T. Braine}
\author{R. Cervantes}
 \author{N. Crisosto}
\author{N. Du}%
\author{G. Leum}
\author{P. Mohapatra}
  \author{T. Nitta}
 \author{L. J Rosenberg}
  \author{G. Rybka}
   \author{J.~Yang}
  \affiliation{University of Washington, Seattle, Washington 98195, USA}
   
\author{John Clarke}
\author{I. Siddiqi}
  \affiliation{University of California, Berkeley, California 94720, USA}

\author{A. Agrawal}
\author{A. V. Dixit}
\affiliation{University of Chicago, Illinois 60637, USA}

\author{M. H. Awida}
 \affiliation{Fermi National Accelerator Laboratory, Batavia, Illinois 60510, USA}
\author{A. S. Chou} 
 \affiliation{Fermi National Accelerator Laboratory, Batavia, Illinois 60510, USA}
 \author{M. Hollister} 
 \affiliation{Fermi National Accelerator Laboratory, Batavia, Illinois 60510, USA}
    \author{S. Knirck}
\author{A. Sonnenschein} 
  \author{W. Wester} 
  \affiliation{Fermi National Accelerator Laboratory, Batavia, Illinois 60510, USA}

\author{J.~R.~Gleason}
  \author{A. T. Hipp}
\author{S. Jois}
 \author{P. Sikivie}
\author{N. S. Sullivan}
\author{D. B. Tanner}
  \affiliation{University of Florida, Gainesville, Florida 32611, USA}

\author{E.~Lentz}
  \affiliation{University of G\"{o}ttingen, G\"{o}ttingen 37077, Germany}

\author{R. Khatiwada}
\affiliation{Illinois Institute of Technology, Chicago, Illinois 60616, USA}
\affiliation{Fermi National Accelerator Laboratory, Batavia, Illinois 60510, USA}

\author{G. Carosi}
\author{C. Cisneros}
\author{N. Robertson}
\author{N. Woollett}
\affiliation{Lawrence Livermore National Laboratory, Livermore, California 94550, USA}

\author{L. D. Duffy}
  \affiliation{Los Alamos National Laboratory, Los Alamos, New Mexico 87545, USA}

\author{C. Boutan}
\author{M. Jones}
\author{B. H. LaRoque}
\author{N. S.~Oblath}
\author{M.~S. Taubman}
  \affiliation{Pacific Northwest National Laboratory, Richland, Washington 99354, USA}

\author{E. J. Daw}
  \author{M. G. Perry}
  \affiliation{University of Sheffield, Sheffield S10 2TN, UK}
\author{J. H. Buckley}
\author{C. Gaikwad}
\author{J. Hoffman}
\author{K. Murch}
  \affiliation{Washington University, St. Louis, Missouri 63130, USA}
  \author{M. Goryachev}
\author{B. T. McAllister}
\author{A. Quiskamp}
\author{C. Thomson}
\author{M. E. Tobar}
\affiliation{University of Western Australia, Perth, Western Australia 6009, Australia}
 
\collaboration{ADMX Collaboration}\noaffiliation

\author{V. Bolkhovsky}
\affiliation{MIT Lincoln Laboratory, 244 Wood Street, Lexington, MA 02421}
\author{G. Calusine}
\affiliation{MIT Lincoln Laboratory, 244 Wood Street, Lexington, MA 02421}
\author{W. Oliver}
\affiliation{MIT Lincoln Laboratory, 244 Wood Street, Lexington, MA 02421}
\affiliation{Massachusetts Institute of Technology, 77 Massachusetts Avenue, Cambridge, MA 02139}
\author{K. Serniak}
\affiliation{MIT Lincoln Laboratory, 244 Wood Street, Lexington, MA 02421}

 
\date{\today}
\begin{abstract}
We present a new exclusion bound of axion-like particle dark matter with axion-photon couplings above $\mathrm{10^{-13}}$ $\mathrm{GeV^{-1}}$ over the frequency range 4796.7--4799.5 MHz, corresponding to a narrow range of axion masses centered around 19.84 $\si\micro$eV.
This measurement represents the first implementation of a Josephson Traveling Wave Parametric Amplifier (JTWPA) in a dark matter search. The JTWPA was operated in the insert of the Axion Dark Matter eXperiment (ADMX) as part of an independent receiver chain that was attached to a 0.588-liter cavity. The ability of the JTWPA to deliver high gain over a wide (3 GHz) bandwidth has engendered interest from those aiming to perform broadband axion searches, a longstanding goal in this field. 
\end{abstract}

\maketitle

Thought to compose 85\%~\cite{Planck} of the matter content of the universe, dark matter appears necessary to explain a variety of observations, from galactic rotation curves and gravitational lensing, to the anisotropies of the cosmic microwave background and the properties of galaxy cluster collisions~\cite{Planck,1980ApJ...238..471R,kaiser1993mapping,robertson2016does}. Despite the scope of these observations, the nature of dark matter remains a mystery. One particularly promising candidate is the axion, which has the unique ability not only to account for all the dark matter~\cite{Preskill:1982cy,Abbottf:1982a,DINE1983137}, but also to solve the so-called strong-CP problem~\cite{Peccei1977June,Peccei1977Sept,Weinberg:1977ma,Wilczek:1977pj}.

One attempt to render the axion visible is called the resonant cavity haloscope, first proposed by Pierre Sikivie~\cite{Sikivie1985,Sikivie:1983ip}. The Axion Dark Matter eXperiment (ADMX) is one such experiment that is unique in its ability to detect Dine-Fischler-Srednicki-Zhitnitsky (DFSZ) axions~\cite{Dine:1981rt,Zhitnitsky:1980tq,bartram2021axion}. Other haloscopes have achieved sensitivity to the Kim-Schifman-Vainshtein-Zakharov (KSVZ) model of the axion~\cite{Kim:1979if,Shifman:1979if,PhysRevD.97.092001,Brubaker:2017rna,PhysRevLett.118.061302,PhysRevLett.80.2043,PhysRevD.69.011101,1538-4357-571-1-L27}. Of the two, the DFSZ axion is the more difficult to detect, with signals an order of magnitude smaller than the KSVZ axion. Sensitivity to the DFSZ axion is therefore a noteworthy feat. Such progress is attributable to the advent of ultra-low-noise quantum amplifiers such as the Microstrip SQUID Amplifier (MSA)~\cite{muck1998radio} and the Josephson Parametric Amplifier (JPA)~\cite{PhysRevB.83.134501,PhysRevApplied.8.054030}. The subsequently developed Josephson Traveling Wave Parametric Amplifier (JTWPA)~\cite{JoseAumentado} provides power gain over a much wider bandwidth, which could enable broadband axion searches.


The JTWPA consists of a lumped element transmission line, where Josephson junctions are the non-linear inductive element. When a microwave signal travels down the line, the non-linear inductance causes four-wave mixing. This feature enables the JTWPA to provide broadband power gain~\cite{JoseAumentado}. Due to this feature, it is recognized that the JTWPA is generally well-suited for axion searches that aim to cover a wider range of the axion parameter space. Broadband experiments are ideal for axion searches involving transient signals appearing off-resonance, and allow for the possibility of multimode searches~\cite{arvanitaki2020large,rosenberg2015final}. Another advantage of the JTWPA is that it does not require a circulator to separate incoming and outgoing modes, enabling a compact receiver design. This helpful feature is ideal for axion searches hoping to take advantage of any extra space in a volume that is typically constrained by the diameter of a magnet bore.

\begin{figure}
    \centering
    \includegraphics[width=0.5\textwidth]{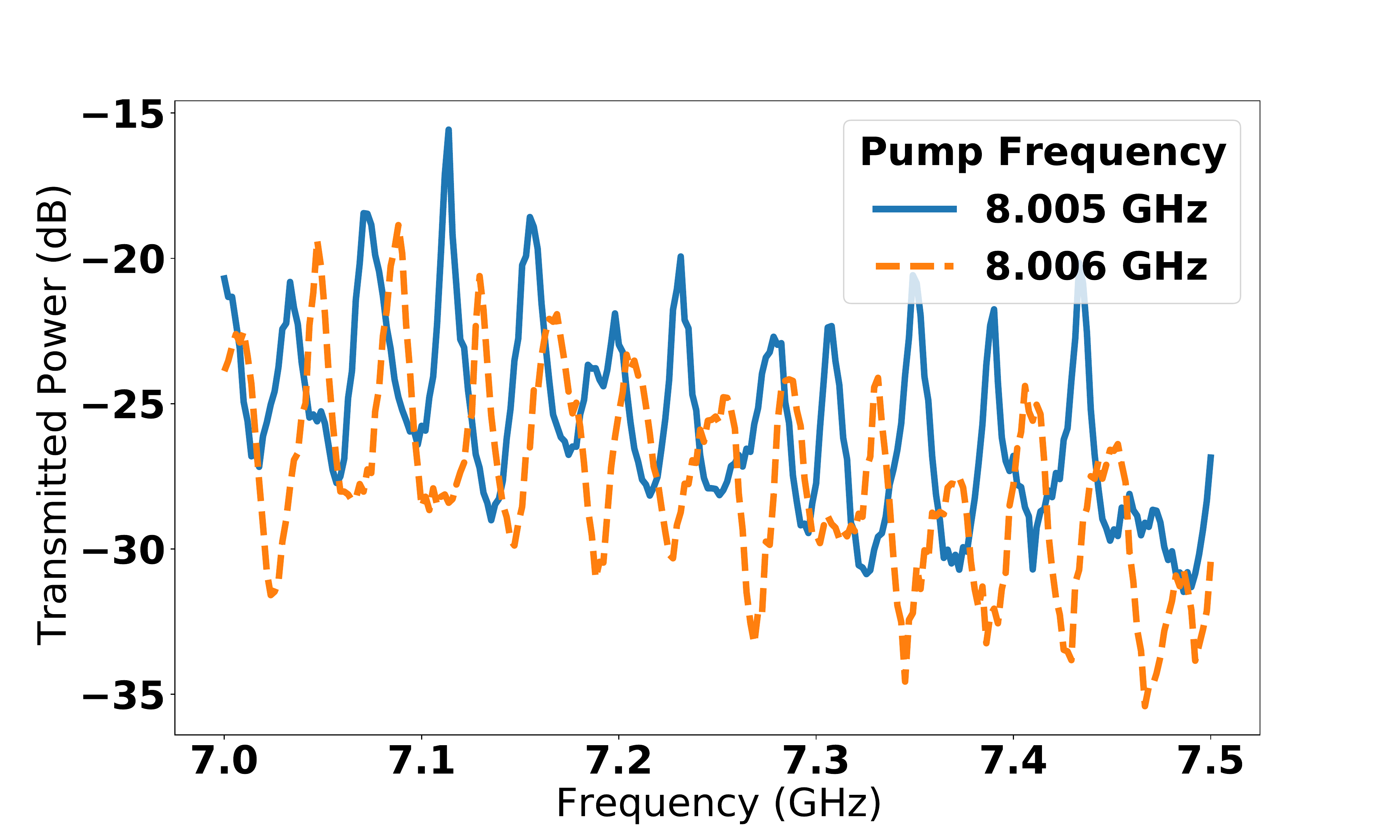}
    \caption{Reflection measurements, acquired by routing power from the cavity bypass line through the output line, with two different pump frequencies applied to the JTWPA, shown in solid (blue), and dashed (orange) curves. The gain peaks are spaced roughly 38 MHz apart. The JTWPA was biased such that the 2-MHz power spectrum was centered on a region with maximum gain.}
    \label{fig:twpa_refl}
\end{figure}

The wide bandwidth of the JTWPA simplifies the amplifier optimization process. A JTWPA requires only the adjustment of the pump frequency and power, whereas a JPA requires the adjustment of the former plus a bias current to tune its resonant frequency. A feature of the JTWPA is that its gain is not flat across a wide range. Instead, there are oscillations in the gain by about 10 dB, with peaks spaced roughly 38 MHz apart. This feature arises from imperfect impedance matching to the Josephson Junctions in the transmission line that constitutes the JTWPA. The exact spacing and height of the peaks is dependent on the individual JTWPA. Adjusting the frequency of the JTWPA pump tone can adjust the frequencies of these peaks, and therefore optimize the gain at a particular frequency. A demonstration of this effect is shown in the plot of the receiver gain as a function of frequency, shown in Fig.~\ref{fig:twpa_refl}. 

\begin{figure}
    \centering
    \includegraphics[width=0.4\textwidth]{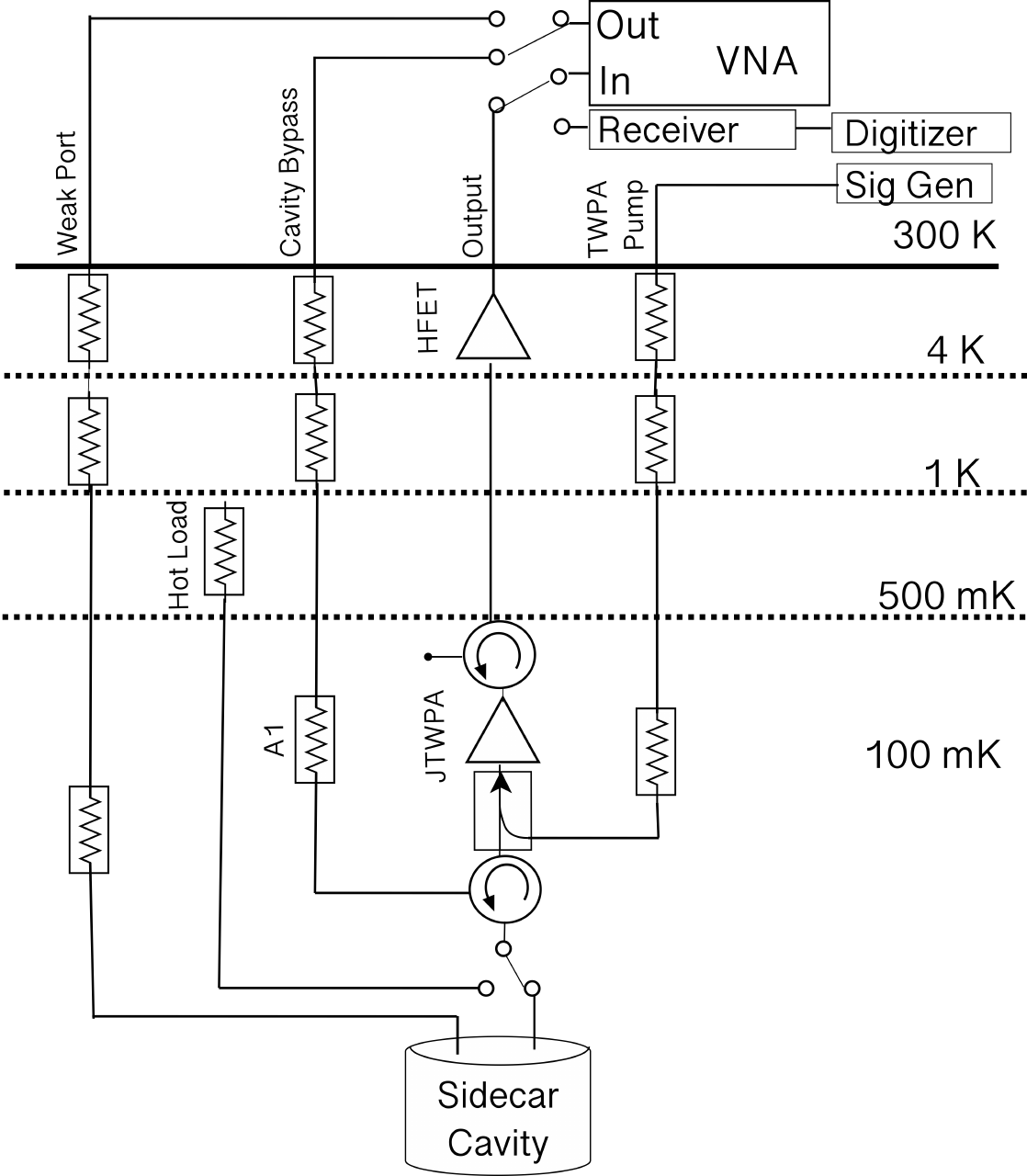}
    \caption{RF schematic of the sidecar cavity for this run. The JTWPA resides between two circulators on the output line. These two circulators are used to prevent power from reflecting back into the cavity. All these components are mounted within a so-called quantum amplifier package that is thermally sunk and bolted to the main ADMX cavity via a coldfinger. The package itself resides in a low-field region that is maintained using a field cancellation coil.}
    \label{fig:recvrchain}
\end{figure}

In an effort to integrate a JTWPA into an axion search, a prototype experiment was operated within the ADMX magnet bore. The ``sidecar'' cavity is mounted on top of the ADMX main cavity~\cite{bartram2021axion}, residing in a magnetic field of about 3.81 T just above the main experiment's superconducting magnet~\cite{thesisBoutan}. Although it uses an independent receiver chain, its operation is secondary to the main experiment. Data-taking is therefore subject to stop when technical issues pertaining to the main experiment are encountered. The 0.588-$\ell$ sidecar cavity can be disassembled into two halves, thus allowing a single tuning rod to be mounted inside with ease. Typical measurements of the sidecar quality factor, or $Q$, were approximately 700. The quality factor was lower than expected due to RF leakage attributable to poor alignment of the two cavity halves. The cavity is coupled by an antenna to a receiver chain that exists independently of the main experiment. At present, the current iteration of the sidecar uses a JTWPA as its first-stage amplifier and a heterostructure field-effect transistor (HFET) amplifier as its second-stage amplifier~\cite{SCHFET}. A diagram of the receiver chain for the sidecar cavity can be seen in Fig.~\ref{fig:recvrchain}. The receiver chain allows for transmission and reflection measurements to determine the cavity frequency, quality factor, and antenna coupling. The JTWPA is located on the output line, acting as the first stage amplifier and conduit for power coming from the cavity. The JTWPA is operated by providing a pump tone coming from a local oscillator in the warm electronics space. The particular JTWPA used in this search was fabricated at and acquired from Lincoln Labs~\cite{TWPA2}. It is made from niobium, with a critical temperature of 9.3 K. Data were acquired for about two weeks with the sidecar cavity until the main experiment required a magnet rampdown. Our first implementation tested only a narrow frequency range that was accessible with the sidecar cavity due to a sticking tuning rod that required repair after completion of the main experiments operation. Nevertheless, this represents the first axion search with a JTWPA. 

\begin{figure}
    \centering
    \includegraphics[width=0.5\textwidth]{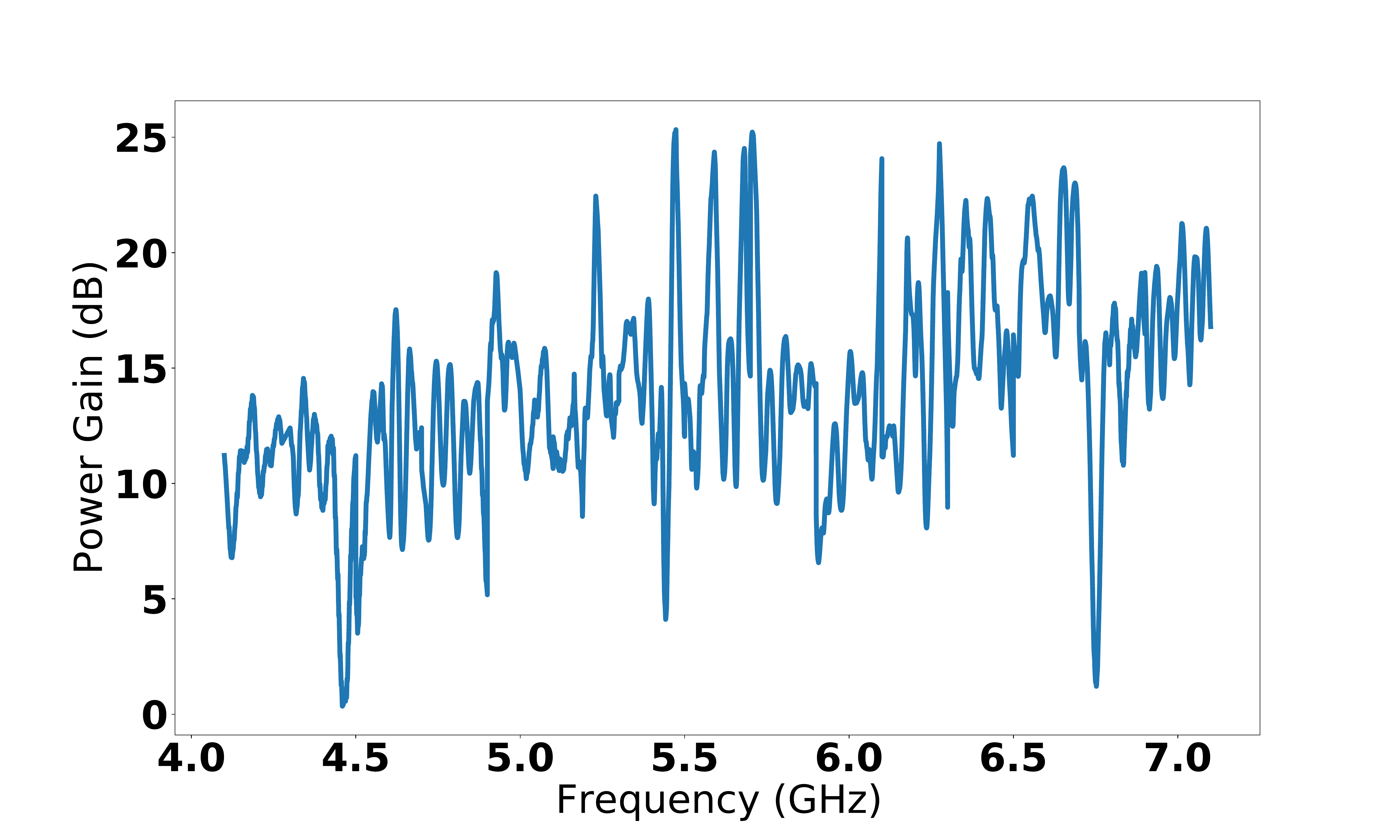}
    \caption{Wideband power gain of the JTWPA from 4-7 GHz, measured during data-taking operations.}
    \label{fig:twpagain}
\end{figure}

Throughout the course of data-taking, the JTWPA gain and signal-to-noise ratio improvement (SNRI) were monitored and rebiased. Occasional rebiases were necessary because of changing temperature conditions and mechanical vibrations of the cavity and receiver chain. Therefore, the JTWPA pump frequency and the local oscillator frequency were adjusted as the frequency shifted. A JTWPA SNRI greater than 9 dB was maintained over the course of 2 weeks, as well as a measured power gain between 12 and 17 dB over the same time frame, at the frequency of the $TM_{\mathrm{010}}$ mode. This power gain was evident over the range of several GHz. Figure~\ref{fig:twpagain} shows the gain as measured over a 3-GHz range, with the characteristic peaks and troughs. 

To perform an axion search with the sidecar cavity, it is necessary to understand the system noise temperature, which can be written as
\begin{equation}
    T_{\mathrm{sys}}=\frac{T_{\mathrm{HFET}}}{{\epsilon}\,\mathrm{SNRI}}.
\label{eq:tsys}
\end{equation}
Here, $T_{\mathrm{HFET}}$ is the noise temperature of the HFET and downstream warm receiver components, $\epsilon{\textless}1$ is the transmission efficiency between the JTWPA and the cavity, and $\mathrm{SNRI}$ is the signal-to-noise ratio improvement when the JTWPA is turned on.
The value of $T_{\mathrm{HFET}}$ was measured during a period in which the experiment was warmed from about 0.2 K to 0.6 K. The hot load, as shown in Fig.~\ref{fig:recvrchain}, was not used due to an unreliable RF switch. In this scenario, the power measured off cavity resonance is defined by
\begin{equation}
P=G_{\text{HFET}}k_{\text{B}}b\left[T_{\text{attn}}\epsilon+T_{\text{cav}}(1-\epsilon)+T_{\text{HFET}}\right],
\label{eqn:power_fit_equation}
\end{equation}
where $T_{\mathrm{attn}}$ is the physical temperature from the final stage attenuator, labelled as $\mathrm{A_1}$ in Fig.~\ref{fig:recvrchain}, on the sidecar bypass RF line, $T_{\mathrm{cav}}$ is the physical temperature of the lines leading from the cavity, and $b$ is the bandwidth.

\begin{figure*}[!htb]
    \centering
    \begin{minipage}{0.5\textwidth}
        \centering
        \includegraphics[width=1.0\textwidth]{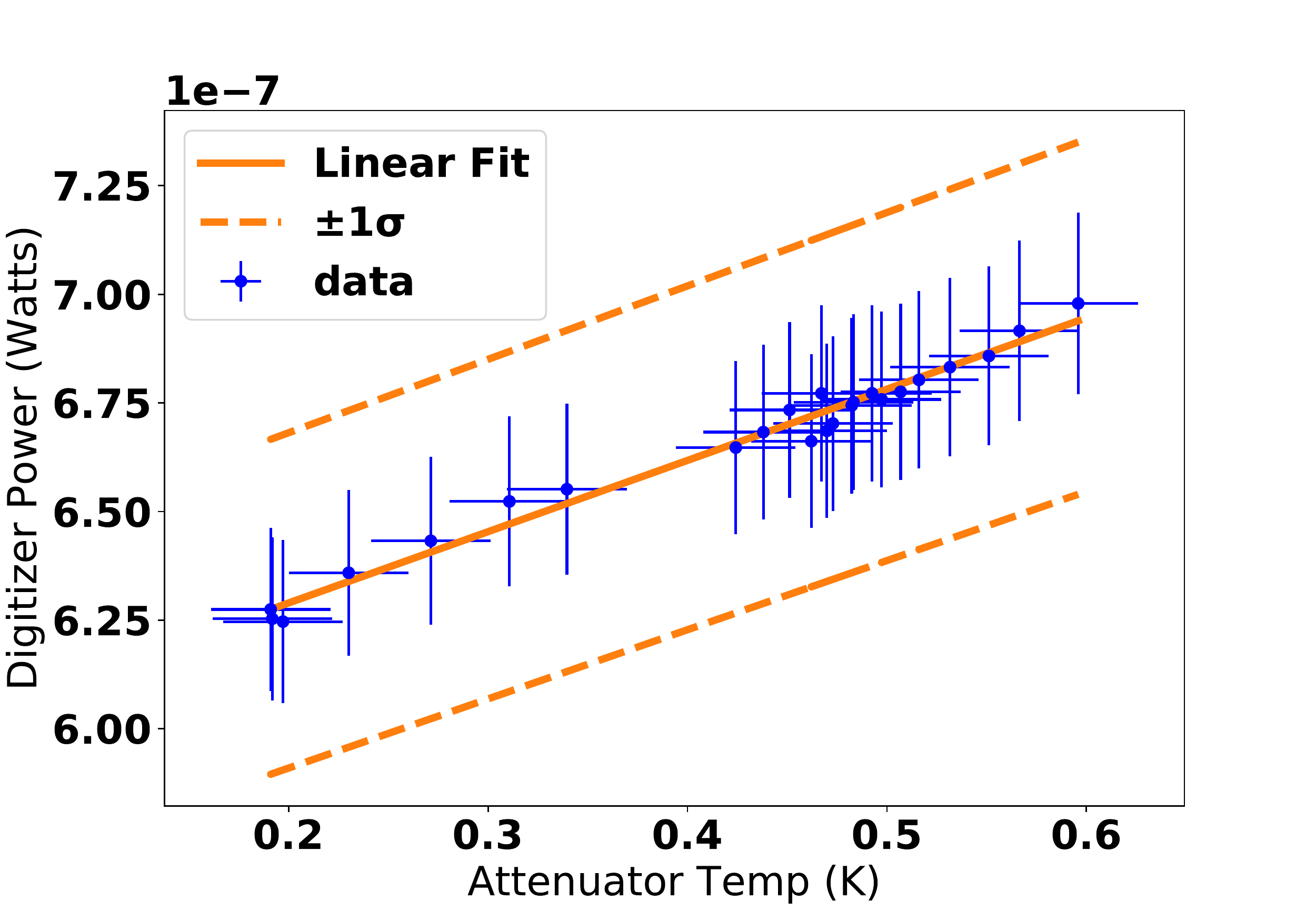}
        \label{fig:prob1_6_2}
    \end{minipage}%
    \begin{minipage}{0.5\textwidth}
        \centering
        \includegraphics[width=1.0\textwidth]{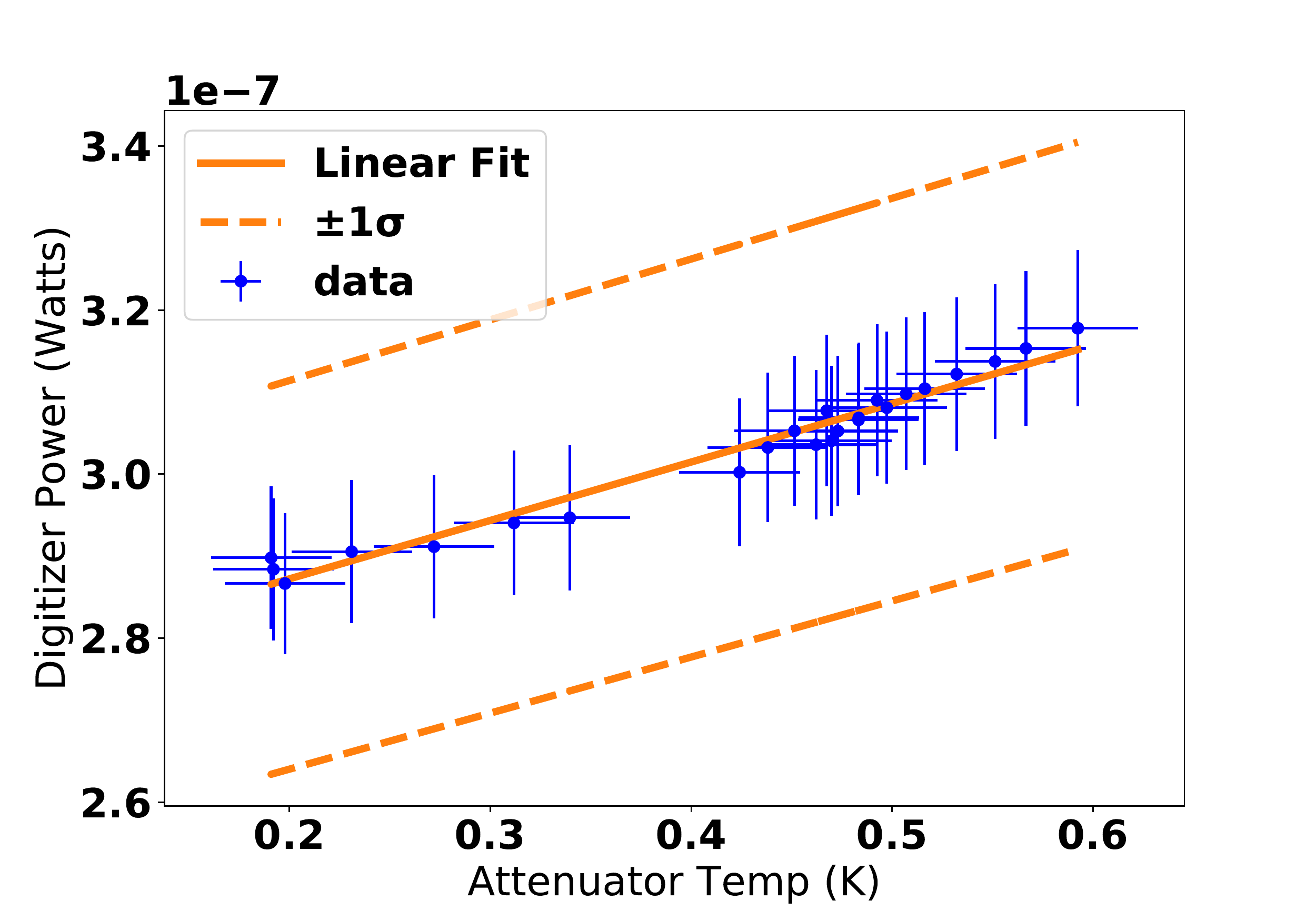}
        \label{fig:prob1_6_1}
    \end{minipage}
    \caption{Mean digitized power of sidecar as a function of temperature of the final stage attenuator for two frequencies off-resonance: 4.6 GHz (left) and 4.9 GHz (right). The actual temperature sensor for the $x$-axis was mounted to the quantum amplifier package, which is thermally connected to the final stage attenuator on the bypass line. The gain of the second-stage and downstream electronics and their associated noise temperature were extracted from the fit. Each data point consisted of an integration time of 10 s. Data are not evenly spaced in temperature-space because of non-linear warming of the ADMX insert. To compute this receiver noise temperature at the resonant frequency, we interpolated between these two values of the receiver noise temperature. The receiver noise temperature, along with the JTWPA SNRI, were used to compute the system noise. The 1$\sigma$ error bounds come from taking the square root of the diagonal of the covariance matrix.}
     \label{fig:hotloadfit}
\end{figure*}

Our measurement of the receiver noise was acquired during the period over which the experiment was being warmed. Because the final stage attenuator is thermally sunk to the main cavity along with the sidecar cavity, Eq.~\ref{eqn:power_fit_equation} then reduces to
\begin{equation}
P=G_{\text{HFET}}k_{\text{B}}b\left[T_{\text{attn}}+T_{\text{HFET}}\right].
\label{eq:powerhfet}
\end{equation}
During the warm-up, power from the receiver was sampled and the temperature of the attenuator was measured. A plot of the power as a function of the attenuator temperature, and the associated fit to the data, can be seen in Fig.~\ref{fig:hotloadfit}. The fit parameter, $T_{\mathrm{HFET}}$, was measured to be 3.7$\pm$0.2 K at a frequency of 4.798 GHz. 

The system noise was computed using Eq.~\ref{eq:tsys}, with this value for $T_{\mathrm{HFET}}$. The SNRI was computed by smoothing and interpolating between measurements of the JTWPA SNRI before and after sampling power from the cavity to acquire a reasonable value over the time duration of data acquisition. 

\begin{figure}
    \centering
    \includegraphics[width=0.45\textwidth]{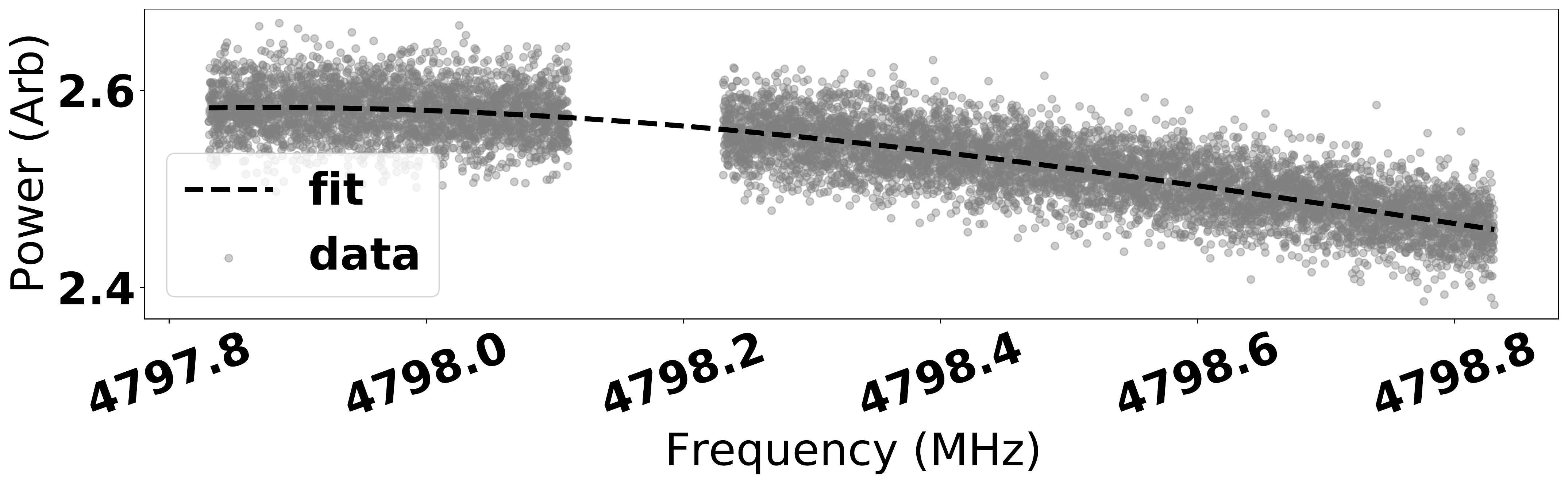}
   \caption{Padé approximant fit to an individual raw spectrum for sidecar. A strong peak in the spectrum was determined to be interference coming from IF electronics, as it appeared always in the same bins, even in power spectra acquired far off cavity resonance. It was removed by masking bins 2800-4000, explaining the gap in the data, above.}
    \label{fig:padefit}
\end{figure}

\begin{figure}
    \centering
    \includegraphics[width=0.5\textwidth]{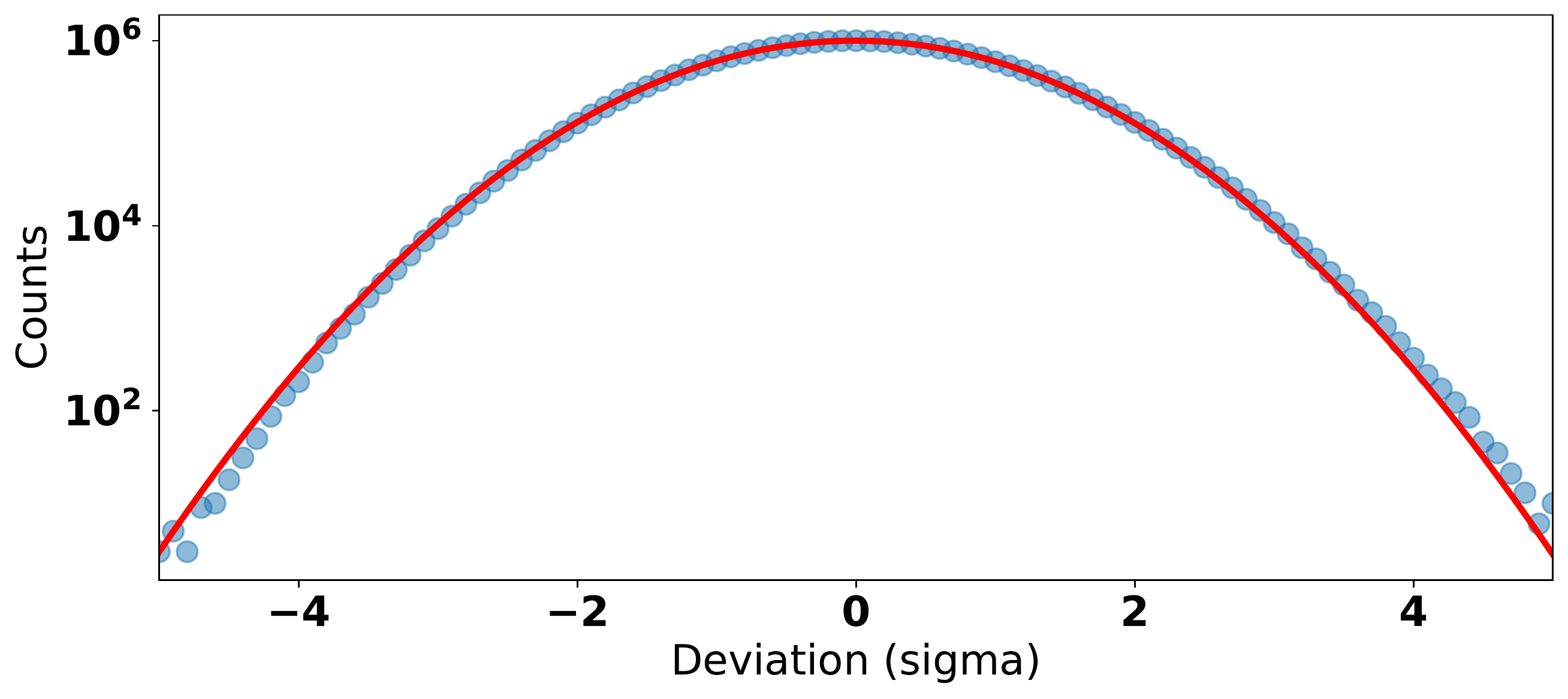}
    \caption{Noise distribution (blue) and Gaussian fit (red) of the data after removing the receiver shape for sidecar. After background subtraction, the distribution of powers measured is well represented by a Gaussian, as expected for white noise. The mean of the fit was computed to be -0.007, with a sigma of 0.990.}
    \label{fig:gauss}
\end{figure}

We measured the attenuation of all cables between the cavity and the JTWPA in liquid nitrogen. We then scaled the measured attenuation to 100 mK, the temperature stage of the quantum amplifier package, using the temperature scaling ratios described in the cable specifications sheet~\cite{Keycom}. The value of $\epsilon$, which we measured to be 0.40$\pm$0.04 in linear units, was then used in the computation of the system noise. This measurement includes attenuation from the JTWPA, which has an associated insertion loss at 4.798 GHz of -3.0$\pm$0.3 dB~\cite{macklin2015near}. Assuming the average JTWPA SNRI of 11.25 dB, the system noise was measured to be 925$\pm$80 mK, with the dominant uncertainty coming from the uncertainty in the attenuation from the cavity to the JTWPA. This value is within the expected range of the system noise using a JTWPA~\cite{simbierowicz2021characterizing}.
\indent Raw data from the sidecar digitizer consisted of 2-MHz wide power spectra with 20,000 bins, each 100 Hz wide. The first and last 5,000 bins were removed from all spectra because the cavity Lorentzian shape was distorted due to intrinsic distortions in the gain of the JTWPA. Bins at the edge of the Lorentzian are least valuable to the data as they exist far from resonance. In a truly broadband experiment that was not limited by the bandwidth of the cavity and receiver electronics, these distortions could be removed, enabling a wide frequency range to be probed. Individual power spectra from sidecar were processed by first dividing by the measured warm electronics transfer function and then applying a Padé approximant filter to remove the transfer function of the cold receiver chain. To acquire the warm electronics transfer function, we terminated the input of the warm receiver, then sampled the power from the warm receiver. An example of the Padé approximant fit (accounting for the warm receiver shape) to the raw spectrum can be seen in Fig.~\ref{fig:padefit}. Removing the transfer functions of the warm and cold electronics resulted in power spectra having the expected Gaussian noise distribution. See Fig.~\ref{fig:gauss}. A strong peak between 4797.5 and 4797.75 MHz was determined to be due to radio-frequency interference by examining scans far from cavity resonance and observing that the peak occupied the same IF bins of the power spectrum regardless of the cavity frequency. A mask was created to exclude these frequencies from the final limit plot. The total bandwidth that was excluded from any individual raw spectrum due to the mask corresponded to 120 kHz.

Axion candidates were defined as any power excesses above 3$\sigma$ fluctuation above the average power level from the flattened spectrum. Two candidates were flagged, but were not identified as axion-like due to lack of persistence between scans. The systematic uncertainties are shown in Table~\ref{tab:uncertainties}. The dominant systematic uncertainties come from the attenuation in the RF cable that runs from the cavity to the JTWPA and the SNRI fit.
\begin{table}
\centering
\renewcommand{\arraystretch}{1.3} 
\begin{ruledtabular} 
\begin{tabular}{@{}lc@{}}
     Source & Fractional uncertainty  \\
     \toprule[0.1ex]
     \hline
     $B^2VC_{\text{010}}$ & 0.04 \\
     $Q$ & 0.2 \\
     Antenna Coupling & 0.01 \\
     $T_{\mathrm{HFET}}/\epsilon$ & 0.11 \\
     SNRI measurement & 0.11 \\
     \bottomrule[0.1ex]
     \hline
     Total on power & 0.26 \\
     \end{tabular}
\end{ruledtabular}
\caption{Dominant sources of systematic uncertainty. The uncertainties were added in quadrature to attain the uncertainty on the total axion power from the cavity, shown in the bottom row. For the first entry, $B$ is the magnetic field, $V$ is the volume, and $C_{\text{010}}$ is the form factor. The last row shows the total uncertainty on the axion power from the cavity.}
\label{tab:uncertainties}
\end{table}
\begin{figure*}
    \centering
    \includegraphics[width=0.9\textwidth]{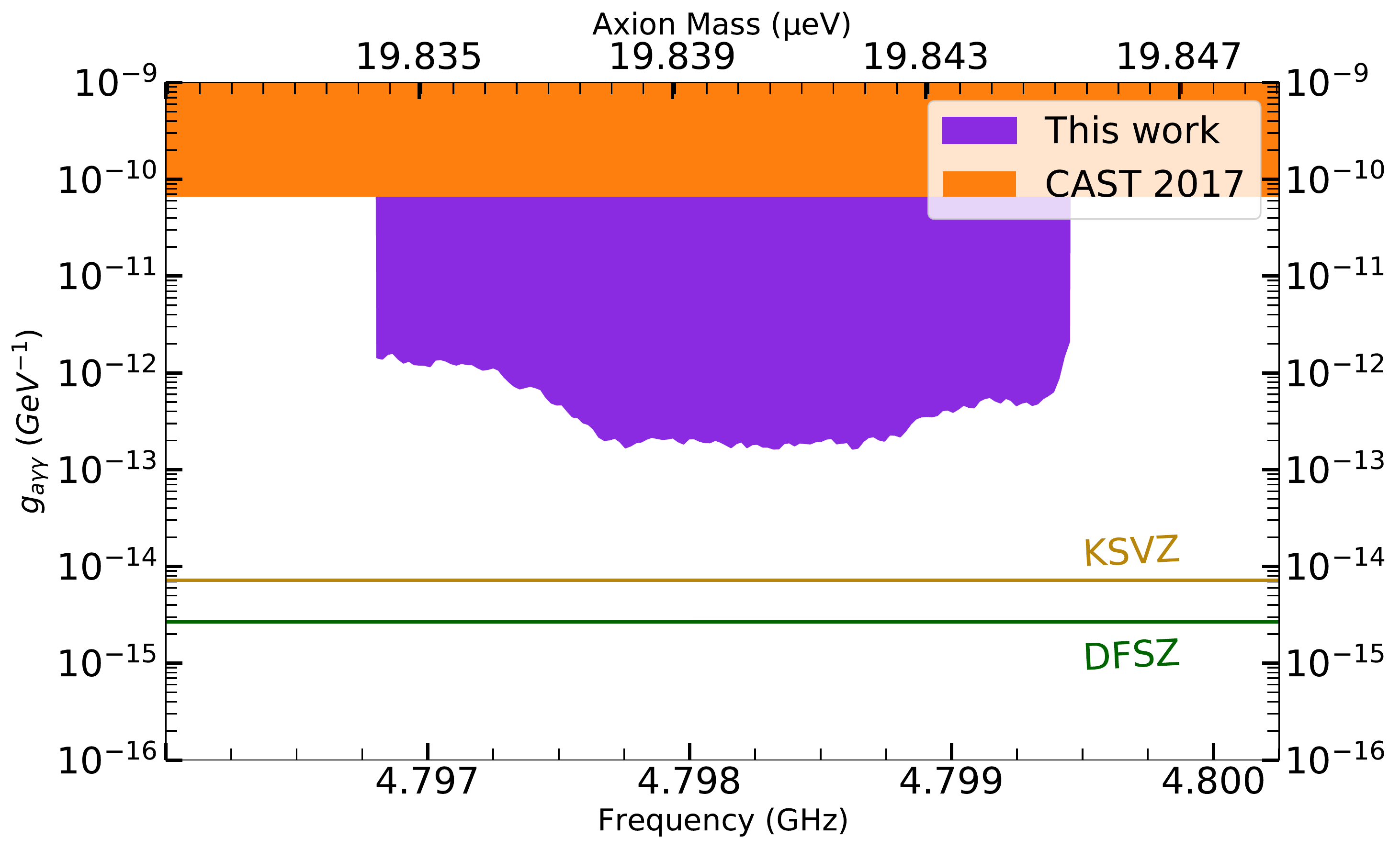}
    \caption{Limit plot from the sidecar data with a JTWPA, assuming  an axion density of 0.45 GeV/cc, equal to the total dark matter density at the Earth. The two lines represent DFSZ and KSVZ models.}
    \label{fig:limit}
\end{figure*}
A limit was set around an axion mass of 19.84 $\si\micro$eV, shown in Fig.~\ref{fig:limit}, which excludes axion dark matter with an assumption that axions constitute all of the dark matter, taken to have a density of 0.45 GeV/cc. Optimal filtering was used with the assumption of a Maxwell-Boltzmann line shape~\cite{turner1990periodic}. The shape of the excluded region shown in Fig.~\ref{fig:limit} arises from the fact that the frequency of the $\mathrm{TM_{010}}$ mode shifted during the course of data-taking. These frequency shifts corresponded to filling of the liquid helium reservoir, which is known to induce mechanical vibrations in the insert. The mechanical vibrations cause small shifts in the positions of the antenna and tuning rod. These vibrations resulted in slight shifts in the resonant frequency of the cavity, despite the fact that the cavity was not intentionally tuned during the period of this data acquisition. The resonant frequency of the cavity drifted from about 4.7975 to 4.7990 GHz throughout the course of data-taking.

We have demonstrated the sustained operation of a JTWPA for axion searches. While the data analyzed in this run were acquired in about 2 weeks, the JTWPA provided reasonable gain for a period of several months during engineering runs prior to data acquisition. Furthermore, an incident occurred in which an accidental increase in the magnet current led to a magnetic field of about 0.41 T in the vicinity of the JTWPA. As a result, the JTWPA was temporarily disabled, but its operation was fully restored after the insert temperature was raised above the critical temperature of niobium. We note that a similar recovery is expected for both the MSA and JPA. Such resilience against inadvertently applied fields bodes well for future axion experiments. This work sets the stage for broadband axion searches, in which we take full advantage of the ability of the JTWPA to operate with low noise temperature and high gain over a wider bandwidth. Future searches that aim to use a JTWPA should continue to improve the means by which we measure the system noise temperature, and consider including an RF line to bypass the JTWPA. This bypass would provide some ability to isolate the JTWPA and therefore study its impacts on the receiver chain.

\section{Acknowledgements}
This work was supported by the U.S. Department of Energy through Grants No. DE-SC0009800, No. DESC0009723, No. DE-SC0010296, No. DE-SC0010280, No. DE-SC0011665, No. DEFG02-97ER41029, No. DEFG02-96ER40956, No. DEAC52-07NA27344, No. DEC03-76SF00098, and No. DE-SC0017987. Fermilab is a U.S. Department of Energy, Office of Science, HEP User Facility. Fermilab is managed by Fermi Research Alliance, LLC (FRA), acting under Contract No. DE-AC02-07CH11359. Additional support was provided by the Heising-Simons Foundation and by the Lawrence Livermore National Laboratory and Pacific Northwest National Laboratory LDRD offices. LLNL Release No. LLNL-JRNL-825283. UWA was funded by the ARC Centre of Excellence for Engineered Quantum Systems, CE170100009, and Dark Matter Particle Physics, CE200100008. Ben McAllister is funded by the Forrest Research Foundation. Tatsumi Nitta is supported by JSPS Overseas Research Fellowships No. 202060305.

\bibliographystyle{apsrev4-1}
\raggedright
\bibliography{references}
\end{document}